\pdfoutput=1

\documentclass[11pt]{article}
\usepackage[final]{acl}
\usepackage{acl}

\usepackage{times}
\usepackage{latexsym}

\usepackage[T1]{fontenc}

\usepackage[utf8]{inputenc}

\usepackage{microtype}

\usepackage{inconsolata}

\usepackage{tcolorbox} 
\tcbuselibrary{skins,breakable} 

\usepackage{amsmath}

\usepackage{array}    
\usepackage{multirow}    
\usepackage{longtable} 

\title{MixRx: Predicting Drug Combination Interactions with LLMs}

\author{Risha Surana, Cameron Saidock, Hugo Chacon \\
         University of Southern California}

\begin{document}
\maketitle
\begin{abstract}
MixRx uses Large Language Models (LLMs) to classify drug combination interactions as Additive, Synergistic, or Antagonistic, given a multi-drug patient history. We evaluate the performance of 4 models, GPT-2, Mistral Instruct 2.0, and the fine-tuned counterparts. Our results showed a potential for such an application, with the Mistral Instruct 2.0 Fine-Tuned model providing an average accuracy score on standard and perturbed datasets of 81.5\%. This paper aims to further develop an upcoming area of research that evaluates if LLMs can be used for biological prediction tasks.
\end{abstract} 

\section{Introduction}
Motivation for the proposed research stems from the recognition of the limitations inherent in existing one-to-one drug interaction checkers. The National Institutes of Health (NIH) \cite{WinNT} recently published work that marks a significant milestone in this domain: CancerGPT \cite{Li2023-lq}. This research focused on cancer drug cocktails to reveal the potential of LLMs to revolutionize how we approach combination therapies in oncology. However, the scope of such research remains narrowly focused on oncological and rare-tissue applications, leaving a vast landscape of multi-organ, multi-system, and multi-disciplinary areas unexplored.

Our ambition is to extend the efforts of CancerGPT beyond the confines of cancer treatment to encompass a broader range of general medical conditions in an emergency medicine domain. The project will ultimately serve as an assistive tool for first responders and emergency department physicians. This advancement would not only increase the efficiency and confidence with which treatments are prescribed, but also significantly enhance patient outcomes by tailoring drug combinations to the specific needs of individuals, taking into account the multifaceted nature of their conditions.

\subsection{Purpose}
In their exploration of cancer drug cocktails, Li et al. emphasized the importance of understanding the synergy between drugs to infer biological reactions that are not directly observable. They noted that "while structured databases hold a limited amount of biological knowledge, the vast majority resides in free-text literature, which can be utilized to train large language models" (LLMs) for predicting biological outcomes \cite{Li2023-lq}. The team at the NIH supports the use of LLMs, particularly in scenarios "where structured data and sample sizes are limited," proposing these models as "innovative tools for biological prediction tasks" \cite{Li2023-lq}. Furthermore, Li et al. highlighted that the application of LLMs to broader, more complex fields remains an emerging area of research. They expressed an eagerness to expand this research to encompass more general tasks involving multiple organs and systems across various disciplines, specifically targeting Emergency Medicine.

Ultimately, in this new area of research, the initial phase involves assessing how effectively models can determine drug combination synergy. This understanding will enable us to evaluate a model's ability to handle and generalize from messy inputs. Finally, testing the model's capacity to predict unseen reactions will necessitate expert support to ensure accuracy and reliability. In this paper, we complete the first two steps.

\subsection{Existing Methods}
In the experience of the authors, the traditional use of Look-Up Tables (LUTs) has proven to have limitations in the area of drug interaction analysis -- especially in fast-paced medical environments like emergency medicine. These tables are highly dependent on the exactitude of inputs, requiring not only perfectly spelled drug names but also a comprehensive list that includes all possible synonyms, acronyms, and hospital-specific slang. See tables \ref{table:perfect_inputs} and \ref{table:messy_inputs} for experimental LUT results on misspellings and synonym usage. These dependencies make LUTs less adaptable to the often chaotic and hurried inputs typical in emergency settings, where drug names might be misspelled or referred to by general or brand names, descriptions, or even team-specific slang. The interaction checkers based on LUTs are inherently limited; they do not generalize beyond their explicit entries and require constant updates to remain useful, offering 100\% accuracy only with perfect inputs and no room for predicting interactions of unseen drug combinations. Their performance is bounded by the drugs listed and the analyses provided, which could become a bottleneck in a setting where timely decision-making is critical.

\begin{table}[ht]
\centering
\caption{Performance of the Look-Up Table for Perfect Inputs}
\label{table:perfect_inputs}
\textbf{Perfect Inputs Performance}
\par\medskip 
\begin{tabular}{|p{4cm}|p{2cm}|}
\hline
\textbf{Condition} & \textbf{Accuracy} \\
\hline
Known drug combinations & 100\% \\
\hline
\end{tabular}
\end{table}

\begin{table}[ht]
\centering
\caption{Performance of the Look-Up Table for Messy Inputs}
\label{table:messy_inputs}
\textbf{Messy Inputs Performance}
\par\medskip 
\begin{tabular}{|p{2cm}|p{2.5cm}|p{1.5cm}|}
\hline
\textbf{Input Type} & \textbf{Condition} & \textbf{Accuracy} \\
\hline
Synonyms / Acronyms & Exist in PubChem & 100\% \\
\cline{2-3}
 & Do not exist in PubChem & 0\% \\
\hline
Spelling Mistakes & Exact matches & 0\% \\
\cline{2-3}
 & Partial-substring matching & 20\% \\
\hline
\end{tabular}
\end{table}

On the other hand, LLMs present a more dynamic alternative. LLMs are inherently more flexible and can handle messy inputs typical in emergency scenarios, such as misspellings or descriptions. This is a feature that can help reduce Medical Malpractice and the common error if misspelling and confusing drug names for similarly spelled alternatives. Unlike LUTs, LLMs have the capacity to generalize from the data they are trained on, making them capable of predicting drug interactions even for drug combinations that have not been explicitly observed before. They can utilize the extensive information stored in free-text medical literature to provide comprehensive analysis, not limited merely to quantitative data but also contextual insights based on multi-drug histories and patient information. Once trained, these models can be efficiently distributed, reducing the need for continual upkeep. This adaptability and depth of analysis position LLMs as a potentially more effective tool in emergency medical settings, where the variety and complexity of scenarios require a robust and flexible system that can quickly adapt to new information.

\section{Method}
Our codebase can be found at the following Github repository: \cite{repo}.

We evaluate the model on its ability to predict the compatibility of a new drug with existing ones using a defined synergy metric and terminology. This involves assessing if a new medication would synergize well with the patient's current regimen. Second, the project assesses whether the model's predictions are based on sound reasoning, incorporating both broad medical knowledge and specific input data from the patient's drug history. The final critical aspect is the model's ability to recommend a new, additive drug in complex multi-drug scenarios that would positively interact with the existing treatment plan. This final recommendation aspect is left for future work.

\section{Data}
We use 3 datasets in this evaluation:  
1. SynergxDB    
2. SynergxDB, Messy   
3. NIH Drug Encounter Data  

We also build a ground truth dataset used to evaluate our model outputs for accuracy and other metrics during Evaluation.

\subsection{Dataset 1: SynergxDB}
We use SynergxDB \cite{Lab2020-mi}, a database which contains 16,525 synergistic drug-pair combinations for biomarker discovery, each with 4 forms of synergy calculations such as BLISS, LOEWE, HSA, ZIP. These drug-pairs exist based on combinations that often overlap (harmoniously and competitively) in both research and medical practice, so they are a good representation of patient data.

To convert this database into a dataset, we have extracted out the drugs pairs that have existing synergy scores and built a dataset with 10,000 valid drug combinations of randomly varying lengths of 2 to 5 drugs. Our data is split into a 80/10/10 train/test/validation set. For a drug combination list to be considered valid, for each pair of drugs in each combination, there is a corresponding array of synergy scores. This ensures that we can easily verify that the model output is correct using our Python-based synergy validation pipeline that verifies if an answer is correct by checking the 4 synergy scores for each pair of drugs in any given combination.

Using our pre-processing pipeline, our data is then transformed from a tabular input to a string-based model prompt using the following template. 
\begin{tcolorbox}[colback=gray!5!white,colframe=gray!75!black,title=Example Input Prompt,breakable]
\footnotesize
Given the following set of drugs, decide if the synergy of the drug combination is synergistic or antagonistic: Sepantronium Bromide, Crizotinib, Pictilisib.

\texttt{"Pictilisib and Sepantronium Bromide have a Loewe score of: -0.3191, HSA score of: -0.4446, and ZIP score of: -0.7671. Crizotinib and Sepantronium Bromide have a Loewe score of... Generate a prediction on whether the overall drug combination is synergistic or antagonistic. Provide a quantitative synergy score based on the weighted average of Loewe, HSA, and ZIP scores, qualitative reasoning supporting your prediction, and a confidence level in your prediction. Format your answer as specified below:
\{
\textbackslash"Prediction\textbackslash": \textbackslash"Antagonistic\textbackslash",
\textbackslash"Qualitative Reasoning\textbackslash": \textbackslash"This...",
\}
}

\end{tcolorbox}

This format allows us to fine train the model on accurate data relating to drug Additivity, Synergy, and Antagonism, as well as easily and quickly verify our answers.

\subsection{Ground Truth Data}
Our ground truth, to be used in the Evaluation of our model outputs, is generated at this step for each valid drug combination in our dataset. It consists of the calculated prediction label (Synergistic, Antagonistic, or Additive) and a template-generated string reasoning that performs a step-by-step chain-of-thought pairwise analysis. This reasoning is later compared to the reasoning generated by the model in Evaluation. In order to determine the correct prediction label, we performed extensive literature review on the 4 provided synergy scores (Loewe, Bliss, ZIP, and HSA) for each drug pair in the combination. We looked at how they are utilized in lab experiments, what they represent, and the functions to calculate the values.

We then built an algorithm that analyzes the interactions of drug pairs using different scoring methods (Loewe, Bliss, ZIP, and HSA) to classify their synergistic, antagonistic, or additive effects. The function aggregates these classifications to determine an overall interaction result for each pair, takes outlier data into account, and concludes with a final analysis based on the presence of significant synergy, antagonism, and the relative counts of synergistic versus additive interactions.

\subsection{Dataset 2: SynergxDB, Messy}
Our SynergxDB Dataset is easily converted into our "Messy Dataset" using input perturbation modifications to our prompt templates.

Our first modification imitates a spelling error by order. For a given example, one drug listed in the input prompt was modified such that two letters were swapped. This represents the commonly encountered errors of swapping an "i" or "e" while spelling, for instance.

Our second modification imitates a spelling error by silent letters. Here, we removed the last letter of a word. This represents a common mistake of missing the last letter while a medical professional is spelling in a rush, or if there is a silent letter at the end of the word that is not pronounced and hence not remembered in spelling.

We then take our SynergxDB dataset and apply one of the two perturbations to each example, chosen at random. This forms our final Messy Dataset.

\subsection{Dataset 3: NIH Data}
The NIH \cite{WinNT}, using private repositories, provides access to patient symptoms, histories, recommended treatments, and outcomes. We select a specific subset of patient encounters at clinics, ambulances, and ERs to compile our NIH Dataset. This dataset offers a detailed view into the real-life decisions made by doctors, EMTs, paramedics, and other medical professionals in the context of emergency medication. It specifically limits the information to medications used in emergency situations. This data has a size of 73,252 unique medical events, such as emergency room visits and general outpatient visits, that warranted a drug exposure.

We run two experiments with this dataset. First, although no corresponding synergy data is available at the NIH, this limitation allows the model to infer relationships based on factors such as the time of use, duration of use, and side effects. Alternatively, we have cross referenced the drugs in our SynergxDB database to the drugs listed in the NIH data to evaluate the decisions made by real medical professionals. This allows us to get a sense of how synergistic is the average drug combination prescribed by a medical professional, in order to set an upper-bound for model performance. The experiments relevant to this dataset are currently in-progress, and shall serve as the next steps for this project.

\begin{table*}[ht] 
\centering
\caption{Model Output Comparison for Drug Combination Analysis}
\begin{tabular}{|p{0.2\textwidth}|p{0.6\textwidth}|p{0.12\textwidth}|}
\hline
\textbf{\textit{Prompt}} & \textit{\textbf{Context:} BML-190 and Temozolomide have a Loewe score of: -0.0841. The drug combination to analyze is: Temozolomide, BML-190... According to the rule, if Loewe > 0.1, the outcome is Antagonistic; if Loewe < -0.1, it is Synergistic; otherwise, it is Additive... Decide if the combination is synergistic, antagonistic, or additive.} & \textit{Additive} \\
\hline
\textbf{Model} & \textbf{Output} & \textbf{Correctness} \\
\hline
\textbf{GPT-2} & Sent through the Full Disclosure mailing list https://nmap.org/mailman/listinfo/fulldisclosure Web Archives \& RSS: http://seclists.org/fulldisclosure/ By Date By Thread Current thread: BML-190 (B) - (C) 2018-09-27T00:00:00Z - (D) 2018-09-27T00:00:00Z - (E) 2018-09-27T & Incorrect \\
\hline
\textbf{GPT-2 Fine-tuned} & Addition Additives Additionally, Synergy is \textbf{additive}; Symitriptyline and BMS-387064 have synergism scores of -1.0641. Additonib and Symetinib have & Correct \\
\hline
\textbf{Mistral-7B-Instruct-v0.2} & Based on the given Loewe score of -0.0841, the drug combination is \textbf{additive}. Since the Loewe score is less than -0.1 but greater than 0.1, it falls into the additive category. Antagonistic combinations have Loewe scores greater than 0.1, and synergistic combinations have Loewe scores less than -0.1." & Correct \\
\hline
\end{tabular}
\end{table*}

\section{Models}
We use GPT-2 \cite{radford2019language} and Mistral-7B-Instruct-v0.2 \cite{jiang2023mistral} in our experiments. We then finetune both the GPT-2 and Mistral-7B-Instruct-v0.2 model.

Developed by OpenAI, GPT-2 is a large language model with 1.5 billion parameters, pretrained on the WebText dataset derived from text across 45 million website links. As a simple and lightweight model, it serves as an excellent baseline due to its ease of use and the minimal resources required for fine-tuning. This model provides valuable insights into how large language models (LLMs) function and offers a practical understanding of how to effectively prompt a model while managing limitations related to its size, which caps at 1024 tokens including the prompt. However, GPT-2 has its drawbacks; it is not specifically pretrained for medical tasks and often produces incoherent or degenerative outputs, highlighting some of the challenges when using it in specialized domains like healthcare.

MistralAI's model, Mistral-7B-Instruct-v0.2, boasts 7 billion parameters and is specifically fine-tuned for conversation and question answering. As a larger model, comparable to LLama, Mistral is notably well-suited for and easy to fine-tune for a variety of applications, including those that require handling messy inputs. Its ability to be fine-tuned for QA tasks allows it to potentially generalize to the medical field, despite not being pre-trained for medical tasks. However, the larger size of the model means it is more expensive to fine-tune, which could be a consideration for those weighing the costs against the benefits of using this advanced model.

\subsection{Fine-Tuning}
The hardware requirements and performance for running and fine-tuning GPT-2 \cite{radford2019language} and Mistral-7B-Instruct-v0.2 \cite{jiang2023mistral} models vary significantly. For GPT-2 and its fine-tuned version, we utilized NVIDIA RTX A6000 GPUs with CUDA version 10.1. The standard GPT-2 required 48 GB of memory, which matches the capacity of the A6000, and it took approximately 90 minutes for a complete run. The fine-tuned version of GPT-2, despite also needing 48 GB, ran considerably faster, completing training in 45 minutes and additional processes in just 14 minutes.

On the other hand, Mistral-7B-Instruct-v0.2 and its fine-tuned counterpart were handled differently. The base version of Mistral operated on the same NVIDIA RTX A6000 with 48 GB of memory and completed its run in 60 minutes. The Mistral model utilized modifications such as torch.float16 to enhance performance efficiency during its operation. To fine-tune the model, we implemented LORA (Low-Rank Adaptation), a technique that helps optimize the model's adaptability while maintaining a balance between performance and computational efficiency. This approach allowed us to achieve more precise fine-tuning with less memory overhead, thus enhancing the model's ability to handle complex question-answering tasks without significantly increasing runtime. Fine-tuning lasted 120 minutes.

\section{Evaluation Framework}
To assess the effectiveness of MixRx, our evaluation framework is centered on a comprehensive set of key performance metrics alongside some baseline comparisons. These metrics are necessary to accurately and effectively understand the model's capability to predict multi-drug interactions.

The metrics employed to evaluate our model's performance include a focus on both synergy accuracy and synergy reasoning. First, we assess whether the model can accurately predict the effectiveness of a new drug combination using existing drug history and synergy metrics. This involves using Precision, Recall, and F1 Score. Additionally, the model's overall correctness is evaluated through Accuracy, calculated as the ratio of correct predictions to total predictions.

For synergy reasoning, we examine if the model's conclusions are based on valid general knowledge and the input data. Here, text similarity metrics such as ROUGE-1, ROUGE-L, and BLEU are utilized \cite{Lin2004ROUGEAP, Papineni2001-we}. ROUGE-1 measures the overlap of unigram words between the candidate and reference text, while ROUGE-L assesses the longest common sequence of words using longest common sub-sequence statistics, ensuring that the model's reasoning aligns closely with expert human judgments.

Overall, Mistral Fine-Tuned (FT) performed the best across the board.

\begin{table}[ht]
\centering
\caption{SynergxDB Dataset: 1000 drug combination results}
\label{table:result}
\textbf{TP, TN, FP, FN}
\par\medskip 
\begin{tabular}{|l|c|c|}
\hline
\textbf{Metric} & \textbf{GPT-2} & \textbf{GPT-2 FT} \\
\hline
True Positives & 11 & 693 \\
True Negatives & 19 & 0 \\
False Positives & 305 & 2 \\
False Negatives & 1 & 37 \\
Unknown & 664 & 268 \\
\hline
\end{tabular}
\par\medskip 
\begin{tabular}{|l|c|c|}
\hline
\textbf{Metric} & \textbf{Mistral} & \textbf{Mistral FT} \\
\hline
True Positives & 52 & 243 \\
True Negatives & 465 & 579 \\
False Positives & 199 & 16 \\
False Negatives & 145 & 98 \\
Unknown & 140 & 65 \\
\hline
\end{tabular}
\end{table}

\begin{table}[ht]
\centering
\caption{Messy SynergxDB Dataset: 1000 drug combination results}
\label{table:result}
\textbf{TP, TN, FP, FN}
\par\medskip 
\begin{tabular}{|l|c|c|}
\hline
\textbf{Metric} & \textbf{GPT-2} & \textbf{GPT-2 FT} \\
\hline
True Positives & 2 & 442 \\
True Negatives & 49 & 1 \\
False Positives & 887 & 14 \\
False Negatives & 0 & 22 \\
Unknown & 62 & 521 \\
\hline
\end{tabular}
\par\medskip 
\begin{tabular}{|l|c|c|}
\hline
\textbf{Metric} & \textbf{Mistral} & \textbf{Mistral FT} \\
\hline
True Positives & 171 & 14 \\
True Negatives & 31 & 577 \\
False Positives & 580 & 235 \\
False Negatives & 65 & 99 \\
Unknown & 154 & 76 \\
\hline
\end{tabular}
\end{table}

\textbf{Precision and Recall:} We evaluated our models on precision and recall, with Mistral Fine-Tuned (FT) achieving scores of 1.00 and 0.82 on the standard dataset, respectively. Precision measures the reliability of the model in predicting synergistic interactions among drug combinations, indicating that 100\% of the model’s positive predictions were correct, and the recall score suggests that the model identified 82\% of all actual synergistic interactions, showcasing a relatively high effectiveness for our model.

\textbf{F1:} Next, we evaluated our models using F1 score. Mistral FT scored the highest, resulting in a score of 0.90. This is a balanced metric that considers both precision and recall, which is important in the medical domain, where both false positives and false negatives carry significant implications for patient health. Our model's high F1 score indicates it has the capability to balance accuracy and completeness very well.

\begin{table}[ht]
\centering
\caption{SynergxDB Dataset: 1000 drug combination results}
\label{table:result}
\textbf{Precision, Recall, F1}
\par\medskip 
\begin{tabular}{|l|c|c|}
\hline
\textbf{Metric} & \textbf{GPT-2} & \textbf{GPT-2 FT} \\
\hline
Precision & 0.00 & 0.71 \\
Recall & 0.02 & 0.49 \\
F1 & 0.01 & 0.58 \\
\hline
\end{tabular}
\par\medskip 
\begin{tabular}{|l|c|c|}
\hline
\textbf{Metric} & \textbf{Mistral} & \textbf{Mistral FT} \\
\hline
Precision & \textbf{1.00} & \textbf{1.00} \\
Recall & 0.66 & \textbf{0.82} \\
F1 & 0.80 & \textbf{0.90} \\
\hline
\end{tabular}
\end{table}

\begin{table}[ht]
\centering
\caption{Messy SynergxDB Dataset: 1000 drug combination results}
\label{table:result}
\textbf{Precision, Recall, F1}
\par\medskip 
\begin{tabular}{|l|c|c|}
\hline
\textbf{Metric} & \textbf{GPT-2} & \textbf{GPT-2 FT} \\
\hline
Precision & 0.00 & 0.74 \\
Recall & 0.05 & 0.36 \\
F1 & 0.01 & 0.49 \\
\hline
\end{tabular}
\par\medskip 
\begin{tabular}{|l|c|c|}
\hline
\textbf{Metric} & \textbf{Mistral} & \textbf{Mistral FT} \\
\hline
Precision & \textbf{1.00} & \textbf{1.00} \\
Recall & 0.61 & \textbf{0.81} \\
F1 & 0.76 & \textbf{0.90} \\
\hline
\end{tabular}
\end{table}

\textbf{Accuracy:} To check the model outputs, our Python-based validation pipeline utilizes a pairwise evaluation of multi-drug synergy, taking into account existing synergy calculations such as BLISS, LOEWE, HSA, ZIP. The overall accuracy of the Mistral FT model was determined to be 0.82, reflecting the proportion of both synergistic and antagonistic interactions that were correctly predicted. This metric underscores the model's general performance across all predictions but also highlights its success and high reliability.

\begin{table}[ht]
\centering
\caption{SynergxDB Dataset: 1000 drug combination results}
\label{table:result}
\textbf{Accuracy, ROUGE, BLEU}
\par\medskip 
\begin{tabular}{|l|c|c|}
\hline
\textbf{Metric} & \textbf{GPT-2} & \textbf{GPT-2 FT} \\
\hline
Accuracy & 0.02 & 0.49 \\
ROUGE-1 & 0.0111 & 0.0253 \\
ROUGE-L & 0.0111 & 0.0253 \\
BLEU & 0.0502 & 0.0000 \\
\hline
\end{tabular}
\par\medskip 
\begin{tabular}{|l|c|c|}
\hline
\textbf{Metric} & \textbf{Mistral} & \textbf{Mistral FT} \\
\hline
Accuracy & 0.66 & \textbf{0.82} \\
ROUGE-1 & 0.3769 & \textbf{0.5288} \\
ROUGE-L & 0.2721 & \textbf{0.4664} \\
BLEU & 0.0081 & \textbf{0.3614} \\
\hline
\end{tabular}
\end{table}

\begin{table}[ht]
\centering
\caption{Messy SynergxDB Dataset: 1000 drug combination results}
\label{table:result}
\textbf{Accuracy, ROUGE-1, ROUGE-L}
\par\medskip 
\begin{tabular}{|l|c|c|}
\hline
\textbf{Metric} & \textbf{GPT-2} & \textbf{GPT-2 FT} \\
\hline
Accuracy & 0.05 & 0.36 \\
ROUGE-1 & 0.0297 & 0.0165 \\
ROUGE-L & 0.0297 & 0.0165 \\
BLEU & 0.4578 & 0.0000 \\
\hline
\end{tabular}
\par\medskip 
\begin{tabular}{|l|c|c|}
\hline
\textbf{Metric} & \textbf{Mistral} & \textbf{Mistral FT} \\
\hline
Accuracy & 0.61 & \textbf{0.81} \\
ROUGE-1 & 0.3693 & \textbf{0.5175} \\
ROUGE-L & 0.2695 & \textbf{0.4515} \\
BLEU & 0.0116 & \textbf{0.3489} \\
\hline
\end{tabular}
\end{table}

\textbf{ROUGE-1:} We evaluated our model reasoning with ROUGE-1 because it provides a straightforward measure of lexical similarity, crucial for ensuring that the model's output employs relevant terminology. For our evaluation of Mistral FT, we achieved a ROUGE-1 score of 0.5288, which indicates that over half of the words that the model generated were also found in the reference text. This suggests that our model has a good grasp of the relevant vocabulary, though it could still be improved to enhance the exactness of its word usage.

\textbf{ROUGE-L:} We used ROUGE-L to assess the quality of sentence-level structures of our model, ensuring that outputs not only contain accurate terms but also maintain a natural flow. We achieved a score of 0.4664 for Mistral FT, which highlights our model's ability to maintain sequences that are structurally coherent with the reference text. This suggests our model has a decent competency in structuring sentences that logically flow between one another, preserving an essential human-like reasoning in its output.

\textbf{BLEU:} BLEU allowed us to evaluate the accuracy and completeness of our model in generating clinically relevant phrases, which is vital for reliability and usability in practical scenarios. Mistral FT achieved a BLEU score of 0.3614, which demonstrates its capability to recreate specific phrases and expressions that are critical when properly conveying how drugs interact with one another. Our score indicates that the model can reasonably align itself with the reference text verbiage but also shows that we can improve our model's ability to capture longer and more complex sentence structures.

\subsection{Analysis}

\begin{table}[ht]
\centering
\textbf{Model Output Example}
\par\medskip 
\begin{tabular}{|l|c|r|} 
\hline
Field & Model Output \\ \hline
Prediction & Antagonistic \\
Qualitative Reasoning & (see caption) \\
\hline
\end{tabular}
\caption{"This drug combination is classified as antagonistic due to the negative combined synergy score, indicating that the drugs interfere with each other's effectiveness. The Loewe score, which measures the additive effect, and the ZIP score, indicating non-interaction, both support this conclusion. The HSA score, representing the highest single agent, further confirms that the combination is less effective than the most effective drug alone."}
\label{table:example}
\end{table}

Our findings support our hypothesis that a fine-tuned large language model is able to perform well on Drug Interaction tasks. Our fine-tuned Mistral model was able to successfully provide consistent accuracy scores across the messy and unaltered dataset, proving that a LLM is capable of handling the messy inputs expected in an emergency setting, and superior to a look-up table. Further, our accuracy values of 0.815 on average across both datasets for the Mistral fine-tuned model, is close to the average accuracy score from the CancerGPT paper, which ranged from 0.6 to 0.88 \cite{Li2023-lq}. This tells us that this line of work has the potential to predict unseen reactions well, given extensive quality training.

\subsection{Limitations and Qualitative Analysis}
\cite{Yin2014-ix} highlights that positive interactions (i.e. synergistic and additive) among drugs are far more likely than negative ones (i.e. antagonistic). The paper's findings implies that the population prior for positive drug interaction is in the area of 95\% (95\% of 18000 combinations). This poses two considerations. First, a model should theoretically be able to avoid the antagonist label and achieve an accuracy score closer to the priors of the positive classes. Second, what is a good balance among class sample counts for the training set? As to the first consideration, the deviation of our two pre-trained models and GPT2 FT accuracy from that of the positive interaction prior brings up questions of why and how can our models can do better. One reason for the deviation may be that pre-trained models are good at generating tokens from learned distributions, not predicting a class. If so, we may benefit from a set of additional feed-forward (FF) layers with non-linearities trained to aid in classification. The first consideration remains open. On the second consideration, we chose not to balance our training dataset, favoring a sample distribution that is representative of the population as estimated by the work in \cite{Yin2014-ix} and the outcome of our data processing pipeline. In future work, we plan to balance sample count from the three classes and comparison results. The model may better learn the profile of an antagonistic combination when the training set is evenly split, which is an important consideration given the dangers of antagonistic combinations.

\section{Related Work}

\textbf{Cancer drug synergy:} CancerGPT \cite{Li2023-lq} has achieved a measure of success in predicting the synergy of drug pairs in cancer rare tissue with limited sample size. We build on their few-shot learning work and attempt to generalize to synergy predictions among two or more drugs in general settings -- e.g. epilepsy medication.\\
\noindent\textbf{Two and three drug combinations:} We draw on prior drug combination research \cite{Luszczki2021-nf, Luszczki2020-bq} to identify key safety and performance objectives in both synergy and additive drug prediction.

\bibliography{acl_latex}

\end{document}